# Lévy flight search patterns of marine predators not questioned: a reply to Edwards *et al.*

David W. Sims[1,2,3]*, Nicolas E. Humphries[1,4]

[1]*Marine Biological Association of the United Kingdom, The Laboratory, Citadel Hill, Plymouth PL1 2PB, UK*

[2]*Ocean and Earth Science, National Oceanography Centre, University of Southampton, Waterfront Campus, Southampton SO14 3ZH, UK*

[3]*Centre for Biological Sciences, Building 85, University of Southampton, Highfield Campus, Southampton SO17 1BJ, UK*

[4]*Marine Biology and Ecology Research Centre, Marine Institute, School of Marine Sciences and Engineering, University of Plymouth, Drake Circus, Plymouth PL4 8AA, UK.*

*Corresponding author: D.W. Sims (dws@mba.ac.uk)

***Summary:*** **Edwards *et al*. (2012, *PLoS ONE*, vol. 7(10), e45174) question aspects of the methods used in two of our papers published in *Nature* (2008, vol. 451, 1098-1102; 2010, vol. 465, 1066-1069) that report results showing Lévy-walk-like and Lévy-flight movement patterns of marine predators. The criticisms are focused on the applicability of some statistical methodologies used to detect power-law distributions. We reply to the principal criticisms levelled at each of these papers in turn, including our own re-analysis of specific datasets, and find that neither of our paper's conclusions are overturned in any part by the issues raised. Indeed, in addition to the findings of our research reported in these papers there is strong evidence accumulating from studies worldwide that organisms show movements and behaviour consistent with scale-invariant patterns such as Lévy flights.**

1. **Introduction**

Lévy flights are a special class of random walk where displacements (steps) are drawn from a probability distribution with a power-law tail (a Pareto-Lévy distribution). A Lévy flight pattern is composed of many small-step 'walk clusters' interspersed by longer relocations, with this pattern repeated across all scales, such that $P(l_j) \sim l_j^{-\mu}$, with $1 < \mu \leq 3$ where $l_j$ is the flight length (move step length), and $\mu$ the power law exponent. Lévy flights comprise instantaneous steps and therefore have infinite velocities, whereas a Lévy walk refers to a finite velocity walk where displacement is determined after time $t$, which reflects the dynamical process of movement[1]. The interest in Lévy flights (the turning points in a Lévy walk) was extended to biology and ecology from the physical sciences in the 1980s when it was suggested that such a pattern could describe the movements of foraging ants[1]. The potential of Lévy flights as a model probability distribution for exploring complex patterns in biological systems was demonstrated in a landmark study[2] in 1996, in which the foraging movements of wandering albatrosses (*Diomedea exulans*) were shown to conform well to an idealised Lévy flight. Following this study, some of those authors went on to develop a theoretical model of Lévy searching[3] and showed analytically and with simulations that an exponent of $\mu \approx 2$ for the power-law move-step frequency distribution was optimal for encountering sparse and randomly distributed targets, e.g. prey. It was demonstrated that the optimal Lévy flight with $\mu = 2$ resulted in searches that increased the probability of a forager encountering new prey patches. Subsequently, it was proposed[4,5] that because Lévy flights can optimise search efficiencies, natural selection should have led to adaptations for Lévy flight foraging – the so-called Lévy flight foraging (LFF) hypothesis.

The seminal work of Viswanathan *et al.*[2,3] opened the way for many empirical studies on the movement patterns of diverse organisms, from cells to humans, and there have been many studies purporting to show the presence of Lévy flights or Lévy walks in many different species[5]. Over the last few years, however, controversy has emerged as a number of these have been overturned[6,7] because they used small datasets and inappropriate statistical methods for identifying putative power-law behaviour in the move-step-length frequency distributions. The most significant in this respect was the apparent overturning[6] (but see ref. 8) of the original observation[2] of Lévy flights in wandering albatross, where the longest move-steps were wrongly attributed to searching behaviour. This has led some researchers to focus on critically evaluating the statistical methods used to identify power laws and truncated power-laws in movement data. Critically evaluating such studies is both necessary and laudable but needs to be conducted without pre-existing bias to one hypothesis or its alternative.

Since 2007 there have been several studies using large datasets of animal and human trajectories that have presented evidence for the presence of power-law distributed movements. The studies have used ever larger datasets and employed progressively more reliable methodologies as the last few years have progressed, with the most recent investigations in 2012 providing some of the strongest evidence for the presence of scale-invariant behaviours such as Lévy flight patterns in animal movements (e.g. refs. 8, 9). It is against this background that Edwards *et al.*[10] question specific aspects of the statistical model fitting and selection methods that formed a part of two papers that were published by us and our colleagues in *Nature* in 2008 (ref. 11) and 2010 (ref. 12). Here we reply to the principal criticisms levelled at particular parts of the statistical methods used to identify candidate power-laws or truncated power-laws in very large datasets comprising >1 million move steps of free-ranging animal movements from multiple species recorded by electronic tags, a large-scale approach not attempted prior to 2008.

## 2. Results and Discussion

### 2.1 Reply to issues raised about Sims *et al.* "Scaling laws of marine predator search behaviour" *Nature* **451**, 1098-1102 (2008)

One of our principal conclusions in the paper was that "model fits to move-step-length frequency distributions for five species (shark, teleosts, sea turtle, penguin) across diverse taxa closely resembled an inverse-square power law with a heavy tail of increasingly longer steps intermittently distributed within the time series that is typical of ideal Lévy walks" (p.1098, ref. 11). Edwards *et al.*[10] question one method we used in Sims *et al.*[11] to fit model distributions and to compute Akaike Information Criteria (AIC) weights for assessing power law (Lévy) and exponential model best fits to the rank-frequency plot of a subset of bigeye tuna (*Thunnus obesus*) movement data. The key question to be answered is whether what they show overturns our paper's conclusions. In their re-analysis Edwards *et al.*[10] confirm that likelihoods were computed from linear fits of models and by calculating new Akaike weight values conclude no support for the power law compared with the exponential model that was, by contrast, strongly favoured by Akaike weights (*w*AIC=1.0). They suggest that this, in itself, questions our conclusion of Lévy-walk-like behaviour. Our conclusion, however, was formed from analysing 1.2 million move steps from 31 individuals from seven species using four different methods (logarithmic binning with normalisation, maximum likelihood modelling, root mean square fluctuation, and power spectrum analysis). Thus, the current re-analysis does not address directly the majority of our analysis or results reported in Sims *et al.*[11]

The analysis we undertook to test for power-law model fits using Maximum Likelihood Estimation (MLE) and AIC weights was asked for by an anonymous referee during the review process and was included only as Supporting Information (p.1-7 and 16-21; ref. 11). With this method, the likelihood of tested models was calculated from residuals of regression fits which results in erroneous AIC, and the testing of models that do not correspond to valid probability distributions. Prior to 2008 this statistical method was used in many other studies to model power-law probability distributions, although it has rarely been used after that time as more appropriate statistical methods for identifying candidate power-laws have been taken up by biologists and others. We are grateful to Edwards *et al.*[10] for confirming this and in helping to emphasise the most reliable and accurate methods to use when testing for Lévy flights in biological data. Nevertheless, for the specific case of a subset of bigeye tuna data re-analysed by Edwards *et al.*[10], the broad conclusion does not differ from that for bigeye tuna stated in our paper. We showed (p.5 Supplementary Information)[11] that there was virtually no support for the pure power-law fit to that bigeye tuna dataset in a rank-frequency plot. From the plot of the model fits (Fig. S1h)[11] we found that neither power law nor exponential models accounted well for this particular tuna data, especially for data points in the distribution's heavy tail. The presence of a heavy-tailed distribution can be characteristic of a power law (Lévy distribution) so should be represented in model fits if the pattern can be said to be a Lévy pattern rather than a 'Lévy-like' pattern (in which, for example, the step frequency distribution decays slower than a pure power law).

We found that the exponential model, describing an uncorrelated Poisson random process, was not well supported in comparison to an intermediate (quadratic) model that accounted better for the distribution's heavy tail (Fig. S1h)[11]. As we showed, without considering an intermediate model the exponential model was supported, even though it clearly did not fit the tail of the distribution. As we stated in our paper[11], the intermediate model had no statistical or biological justification other than being the simplest (i.e. most parsimonious) alternative model to test the hypothesis that the relationship between rank and frequency is neither strictly linear (power law) nor exponential but resembling a 'mixed' model (for discussion of mixed models see ref. 8). In our paper[11] we were appropriately open about the fact that it could have been concluded that the exponential was a relatively better fit than a power law in some cases simply because other intermediate models were not tested. The purpose of an intermediate model, as stated[11], was to explore more complex and realistic properties of the movement patterns, perhaps comprising elements of both move patterns. That better alternatives may be available to test does not detract from our conclusion that the exponential does not account for the distribution's tail.

This finding is supported by Edwards *et al.*[10] in their Fig. 1A and stated in their results, that plotting both models shows that the exponential decays too fast and thus does not account for the heavy tail any better than the power law fit, despite strong support from Akaike weights (*w*AIC=1.0). This deviation of the tail from an exponential model, where longer steps occur more often than predicted, was one part of our original analysis justifying our conclusion of "Levy-walk-like behaviour", which was an appropriately cautious conclusion at that time. For this bigeye tuna dataset we concluded that the heavy tail was not a Lévy tail but was Lévy-like, that is, longer move steps occurred more frequently than expected by an exponential model. The re-analysis by Edwards *et al.*[10] does not exclude 'Lévy-like' behaviour as an explanation for these longer move steps since the distribution's tail is heavier than the exponential model best fit.

The re-analysis undertaken by Edwards *et al.*[10] is a good example of one deficiency of MLE with *w*AIC when used to fit models to distributions where a reasonable fit to the heavy tail is required. As discussed in our paper (p.3-4, Supplementary Information in Sims et al.[11]), the frequency of longer move steps that make up the heavy tail of a power-law distribution (the right-hand side of the distribution) is low compared with the more frequent smaller steps making up the left hand side and can introduce considerable bias away from fewer data in the distribution's tail. Importantly in this context, MLE model fitting to empirical data plotted as a rank-frequency plot gives equal weight to all points even though the vast majority of points are clustered on the left hand side. This may be a potential problem for model selection in some cases because strong support for a model solely based upon Akaike weight values (e.g. $w$AIC = 1.0; strongest support) may be based on a good fit to the left hand side of the distribution rather than to the majority of the heavy tail also.

To illustrate this point and to indicate the weakness of the conclusions drawn by Edwards et al.[10] about our results we have re-analysed the bigeye tuna data. The bigeye tuna data analysed in Fig. S1h in Sims et al.[11] comprised 29,900 move steps from a single individual that was a subset of the full dataset for this individual (it was one of three tuna datasets analysed, the results of which were shown in Fig. 1c in Sims et al.[11]). We have re-analysed the full dataset for this individual ($n$ = 62,325 steps) following methods given in Clauset et al.[13]. However, we did not fit $x_{min}$ or $x_{max}$ by iterative search since Edwards et al.[10] hold some objections to this procedure (see responses in section 2.2 below for discussion on this point). Instead, $x_{min}$ was fixed at 1 (1 m was the minimum depth resolution of this electronic tag) and $x_{max}$ was fixed at the maximum depth value in the dataset. For this full, individual tuna dataset we found that a truncated power-law model fit was strongly favoured ($w$AIC=1.0) over that of the exponential model ($w$AIC=0.0) even though neither model accounts particularly well for the distribution's heavy tail (Fig. 1, below). This result was also confirmed with goodness-of-fit tests (Fig. 1 legend). What this demonstrates is that using the more robust MLE methods the full bigeye tuna dataset for that individual was best described by $w$AIC model selection as a truncated power-law and not an exponential. Therefore, because the truncated power-law decays too slowly (as does the power law) compared to the tuna move-step distribution, whilst the exponential decays too fast, this confirms our original conclusion of a "Levy-like" movement pattern since the distribution's tail is 'heavier' than that expected by a simple random pattern of movement.

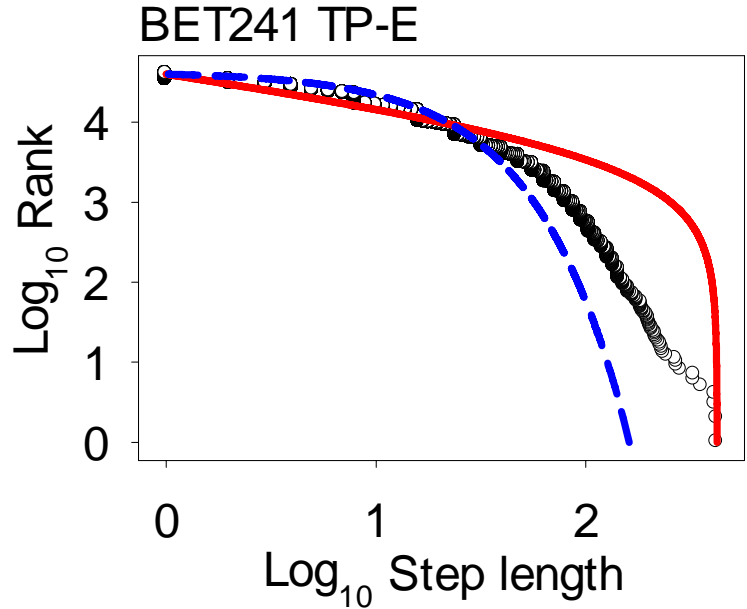

**Figure 1.** Our re-analysis of the full dataset for the individual bigeye tuna ($n$ = 62,325 steps) analysed in Fig. S1h in Sims et al.[11] shows a truncated power-law best fit consistent with a Lévy-like pattern. Best fit truncated power-law (red line) and exponential (blue dashed line) models to observed data (black circles). Truncated Pareto (TP) (power law) distribution was the best fit model compared to the exponential: TP, $\mu$ = 1.366, $x_{min}$ = 1.0 m, $x_{max}$ = 422 m, goodness-of-fit (GOF) = 0.143, $w$AIC = 1.0; exponential fit, $\lambda$ = 0.066, $x_{min}$ = 1.0 m, $x_{max}$ (not fixed), GOF = 0.194, $w$AIC = 0. For full methods used see Humphries et al.[8]

In summary, Edwards et al.[10] question our conclusion of Levy-walk-like movements by marine predators from re-analysis of a subset of data for a single individual (bigeye tuna) comprising <2.5% of the data we analysed and for only one of the methods we used. However, despite the issues Edwards et al.[10] raise, their incomplete re-analysis of the marine predator data in addition to our own re-analysis of the tuna data satisfies us that their findings make no substantive difference to the conclusions reported in Sims et al.[11] Furthermore, since 2008 many papers in this area have used MLE with $w$AIC to test for the presence of power laws in movement data, including those that have analysed very large datasets of movements[8,12,14-16]. These analyses provide strong evidence for both truncated power-laws and exponential movement patterns occurring in diverse species, sometimes with switching between patterns by an individual as it moves from one environment to another. Theoretically, both truncated power-law and exponential movements are expected for animals responding to complex landscapes[3-5,17,18], predictions confirmed in several recent empirical studies showing Lévy patterns of behaviour in individual free-ranging animals[8,12,15].

## 2.2 Reply to issues raised about Humphries *et al*. "Environmental context explains Lévy and Brownian movement patterns of marine predators" *Nature* 465, 1066-1069 (2010)

In this paper we reported movement patterns of 14 species of pelagic marine predator (81 individuals, over 5700 days, totalling $n$ = 12.9 million move-steps) to be well approximated by Lévy flight patterns, in addition to the presence of exponential and more complex 'mixed' patterns. Edwards *et al*.[10] present some opinions about aspects of the method used in Humphries *et al*.[12] where an $x_{max}$ value of a truncated power-law is fitted by an iterative search procedure (that seeks to find the best fit value while maximising the data range included in the best model fit) rather than, as they argue, simply fitting $x_{max}$ to the maximum value in the dataset. As a consequence they argue that key data are ignored. They go on to question whether the data ranges over which the power-law or truncated power-law best fits were found constitute sufficient evidence for power laws. However, we find that these opinions do not stand up to closer scrutiny, either in the light of the results of our quantitative analysis[12] or more recent published literature. We conclude, therefore, that the opinions have no impact on our paper's results or conclusions.

The Lévy flight foraging (LFF) hypothesis tested by us[12] is concerned with detecting scale-invariant movements occurring during foraging behaviour, where walk clusters comprising ratios of different sized steps occur at all scales. Therefore, the presence of a Lévy flight is characterised by the ratios of different-sized step lengths across all scales. Edwards *et al*.[10] state that "the Lévy flight hypothesis is concerned with the rare longer steps in the heavy tail of the data". The implication made by the latter authors is that by fitting an $x_{max}$ below the longest step important move-step lengths may be removed from the analysis. Obviously we are aware that about 10% of data makes up the heavy tail accounting for 90% of a power-law distribution[5]. However in the context of our method, it is hard to see how removing some of the longest steps is problematic because this will reduce the chances of detecting a power law if data points within the heavy tail are removed (as demonstrated by Edwards *et al.*[6] in a previous paper).

It is implied by Edwards *et al.*[10] that without the longest steps being included in our analysis of each individual section, any test is not robust because it is inconsistent with the Lévy flight foraging (LFF) hypothesis. However, this is not a logical conclusion to draw from our procedure: if certain long steps are removed objectively (as the search algorithm does under certain conditions) this is because data points are disregarded that lie outside the scale-invariant structure of a power-law distribution (for example see Fig. 2). In fact, the $x_{max}$-fitting procedure we use in our paper is a very conservative method for testing for a Lévy flight since it may not always include the longest steps; rather it finds the $x_{max}$ value for the largest data range over which a best-fit truncated power-law applies. This is consistent with the LFF hypothesis that tests for the presence of scale-invariant movement structures and should not consider obvious outliers. That we still find truncated power-law best fits after fitting $x_{max}$ confirms to us that the Lévy flight patterns detected are robust and not simply due to a very few points in the heavy tail.

It is most probable that data points (step lengths) were rejected from our $x_{max}$-fitting procedure because they were true outliers that did not form a coherent part of the scale-invariant structure. This is an important consideration for movement time series that will comprise different types of behaviour, from foraging to commuting or social behaviours. Specifically, in the case of large marine predators, such as bigeye tuna, the animals undertake other behaviours that are not part of a foraging pattern and should in fact be excluded from the analysis when testing the LFF hypothesis. Consequently, the fitting of $x_{max}$ to exclude these long movements, which are not part of a best fit power-law distribution, is entirely correct. The analysis performed was to determine whether foraging movements conform to either a Lévy or Brownian pattern, not whether all movements performed by the animal at all times conform to one of these patterns. It is a biological certainty that the animals being studied do not perform the same movement patterns at all times and in our paper[12] we were at pains not only to point this out but, wherever possible, to divide long movement time series into sections that capture bouts of behaviour with more consistency (hence our use of a split-moving window matrix analysis to detect changes in pattern for separating time series into discrete sections). The difficulties with accurately identifying consistent behavioural patterns were explained in our paper[12] and the conclusions drawn were consequently more cautious.

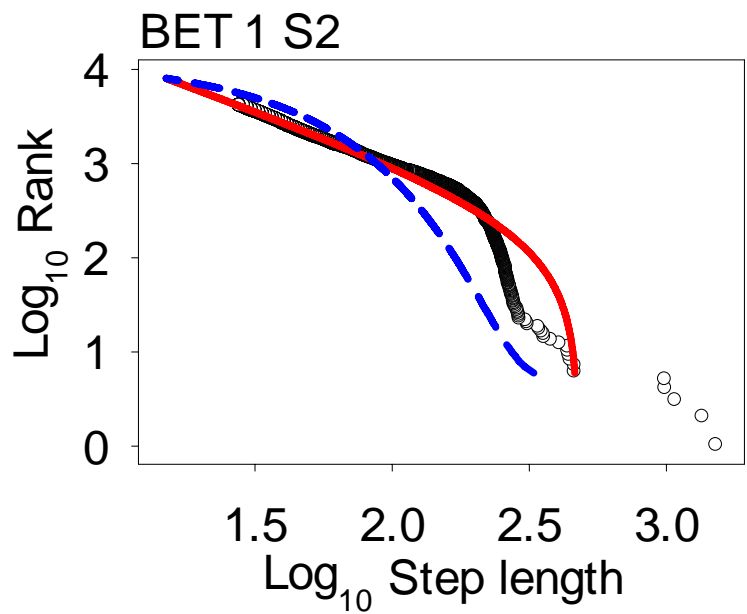

**Figure 2.** The truncated power-law best fit to bigeye tuna (1 section 2) data given in Humphries *et al.*[12] showing how the $x_{max}$-fitting procedure disregards outliers (the 5 longest steps on right) that are well separated from the scale-invariant structure comprising the best fit model. Only 5 move steps were removed from this best fit from a total of 52,806 steps modelled. This method is conservative for detecting truncated power-laws in complex animal movement data.

Nonetheless, the fit of the theoretical distributions to empirical data are not only strongly supported by MLE and *w*AIC but are demonstrably so visually from model fits to empirical step-lengths in rank-frequency plots. Specifically for the case of bigeye tuna 1 (section 2), these longer steps are clearly well separated from the scale-invariant structure making up the best fit model (see Fig. 2) and are likely to be movements associated with behaviours other than foraging, such as deterministic movements associated with behavioural thermoregulation or commuting behaviour. That our method discounts these outliers when they do not form a coherent part of the truncated power-law distribution is consistent with the LFF hypothesis because this hypothesis is concerned with foraging searches and not with other behaviours.

That we used very large movement datasets (with many thousands of move steps analysed per individual) makes the likelihood of mistakenly identifying truncated power-laws greatly reduced by fitting an $x_{max}$ as we did, rather than assuming it to be a particular value (as advocated by Edwards *et al.*[10]). When fitting a truncated power-law it is of course necessary to estimate the value of the $x_{max}$ parameter, otherwise the distribution cannot be fitted. However, to conclude that the correct value for this parameter is always that of the longest step-length recorded is biologically naive, ignoring as it does any possibility of the animal performing other behaviours unconnected with searching. In fact it is equally possible that the true value of $x_{max}$ is larger than the maximum value in the data but was not sampled over the time course of observation. Additionally, Edwards *et al.*[10] do not raise the same criticism about the fitting of $x_{min}$, which has been shown to be important[13]; many very small move-steps captured at the limit of the instrument (here a depth recorder) are likely to be recorded imprecisely and should be discarded and, in any case, it is accepted that power-law behaviour may not cover the full range of movements.

Secondly, in discussing results of our paper Edwards *et al.*[10] imply that by fitting $x_{max}$ we were very selective over the data to which truncated power laws were fitted and that this contributed to the best fits being truncated power-laws as opposed to a competing model, e.g. the exponential. This is an incorrect assumption. Unfortunately however it is not made clear in the Edwards *et al.*[10] critique that in 58% of the movement sections that we found were best fitted by a truncated Pareto (power-law) distribution, the fitted $x_{max}$ value was the same as the maximum move-step length in the dataset. Hence, for 58% of truncated power-law fitted sections our datasets did conform to the method Edwards *et al.*[10] believes to be more consistent with the LFF hypothesis. By not acknowledging this clearly and without bias, Edwards *et al.*[10] misrepresent our paper's results and as such the veracity of their argument is weakened. Furthermore, the $x_{max}$ value is 90% or more of the maximum step length in 73% of the truncated power-law best fitting sections we found. Only in 5 cases is the $x_{max}$ value <50% of the maximum step length and it should be pointed out that these fits still represent best fits over at least 1.5 orders of magnitude of the data. Even if we had chosen to ignore all the sections where the data range fit covered less than 90% of the maximum dataset value, our overall results would be unchanged; truncated Lévy flight patterns would still be prevalent in marine predator movements.

Edwards et al.[10] also assume that when the $x_{max}$ is fitted below the maximum step length in a dataset, a significant amount of data are ignored. To illustrate this potential issue Edwards et al.[10] select to highlight only the example of bigeye tuna 1 (section 2) that has the most extreme difference between the maximum step-length in the dataset (1531 m) and the best fit $x_{max}$ (466 m). However, what is not evident from the example summary data they present is that only 5 move steps (of 52,806 steps in the dataset; i.e. <0.0001%) were excluded from the best fit model by our $x_{max}$ fitting procedure (see Fig. 2). Indeed, as determined by the fitting procedure, these 5 steps also appear upon visual inspection to be true outliers[5] (Fig. 2). Furthermore, for the 19 (out of 60) tracks where the fitted $x_{max}$ is less than the maximum recorded step length we have the following values for excluded step lengths: 209, 100, 44, 27, 17, 14, 11, 7, 7, 6, 5, 4, 3, 2, 1, 1, 1, 1. Therefore, in the majority of cases fewer than 10 steps have been lost from datasets that comprise many tens of thousands of move steps. Thus, this criticism of our method is both incorrect and groundless and does not affect the results we reported[12].

Finally, Edwards et al.[10] question whether the ranges (orders of magnitude) over which truncated power-laws were detected in our free-ranging animal movement data were sufficient to constitute candidate power laws. The latter authors draw attention to a recent proposal of a 'rule of thumb' which suggests that a candidate power law apply over at least two orders of magnitude of the data[19]. Edwards et al.[10] then go on to say that this condition was hardly fulfilled in our paper (only 7 of 66 data sections having bounded power laws were >2 orders of magnitude). In fact 61 sections yielded power-law or truncated power-law exponents within the Lévy range, with 6 sections having exponents <1, so there were 7 sections from 61 (11.5%) showing Lévy fits >2 orders of magnitude of data. However, what Edwards et al.[10] do not state, in what should be a balanced evaluation, is that many more sections were close to this 'rule of thumb': 9 sections (14.8%) had fits over >1.9 orders of magnitude of data, 14 (21.2%) >1.7, and 22 (36%) were >1.5 (for some examples see Fig. 3a-f). Given the difficulty of recording high-temporal-resolution archival data in wild marine predators over sufficient time and space scales to enable a test of the LFF hypothesis, it is perhaps remarkable that >10% of data sections were found to conform to this '>2 orders' criterion of candidate power laws. Figure 3 shows examples of very good fits of marine predator move steps to power law and truncated power-laws, fits which provide very strong evidence supporting the LFF hypothesis, and that should not be ignored. As empirical ecologists we see these as important results (as did the peer reviewers appointed by *Nature*), whereas Edwards *et al.*[10] appear to dismiss this finding, interpret it as a failure of the method and analyses, and suggest this questions the paper's conclusion[12].

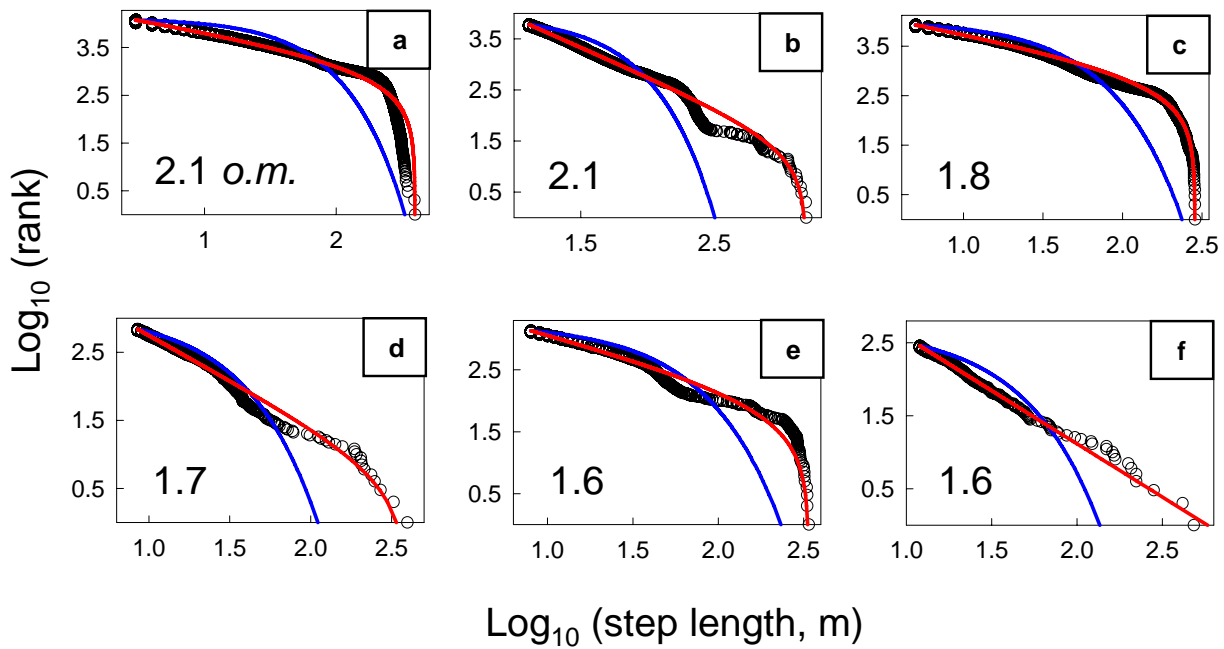

**Figure 3.** Examples from Humphries *et al.*[12] of marine predator best model fits to power-law and truncated power-law (Pareto-Lévy) distributions with fits between 2.1 and 1.6 orders of magnitude (*o.m.*) of the data. Best fit power-law or truncated power-law (red line) and exponential (blue line) models to observed data (black circles). (**a**) Yellowfin tuna 3 section 4; (**b**) bigeye tuna 5 s4; (**c**) yellowfin tuna 1 s3; (**d**) blue shark 12 s2; (**e**) yellowfin tuna 4 s4; (**f**) blue shark 9 s3. Note that the model fit in **f** is to a power law, not a truncated power-law.

Improvements in telemetry technology will allow more comprehensive tests of the LFF hypothesis in the future. In support of this, our paper[12] demonstrates that a higher number of orders of magnitude of data over which a truncated power-law fit occurred correlates with the size of the dataset in terms of the number of data points available for analysis. For example, Table S3 in Humphries *et al.*[12] shows 15 data sections (out of 22) where $x_{max}$ equals the maximum step length and best fits were >1.5 orders of magnitude. This suggests that the technological constraint of tag attachment time (hence, data series length) contributes to a higher frequency of shorter data ranges available for testing the LFF hypothesis, perhaps explaining why most sections were <2 orders of magnitude. Given that the longer step-lengths are rarer it follows that larger datasets are more likely to sample more long step-lengths than smaller datasets. Therefore if longer, larger movement datasets were recorded and exhibited scale-invariant structure, it seems likely that the orders of magnitude over which best fit models apply to data would be increased.

This observation is supported by our recent study[8] published online on 23 April 2012 (but not cited by Edwards *et al.*[10]) that used very high temporal-resolution GPS tracking data (a position each 1 or 10 s) of albatross foraging movements to test for the presence of Lévy flights. The study demonstrated that a significant proportion (31%) of Lévy flight patterns were found among 126 individuals of two species, with 64% of power-law model fits spanning data >2 orders of magnitude, and several >3 orders. This emphasises that better data in the future should allow even more robust and detailed analysis. Moreover, we do not assume that scale-invariant foraging patterns are common in nature and a pattern that should always be present; Lévy flights are theoretically advantageous under certain environmental conditions[5] so would not be expected at all times, and also that, as mentioned previously, animals often engage in behaviours other than searching. What is clear though is that there is strong evidence to support our conclusion of movements approximated by Lévy flights in marine predators.

## 3. Conclusion

In summary, we feel that in making their piecemeal criticisms of Sims *et al.*[11] and Humphries *et al.*[12], Edwards *et al.*[10] have missed a broader conceptual point and in doing so contradict themselves. According to their own notional criteria there were at least seven candidate power-laws in movement data from free-ranging animals reported in Humphries *et al.*[12] that supported the presence of movements approximating a Lévy flight in marine predators. This provides an opposite view to what Edwards *et al.*[10] conclude based on their re-analysis of data presented in Sims *et al.*[11]. In so doing it indicates Edwards *et al.*[10] have presented an incoherent argument in their paper: on the one hand they question the support for Lévy behaviour in marine predators presented in Sims *et al.*[11], but then in criticising Humphries *et al.*[12] they draw attention to truncated power-laws of marine predator movement that have been detected using MLE with *w*AIC and that fulfil the notional rule of thumb on power laws they advocate. This contradiction is self-defeating of their own conclusion[10] that they overall "question the claimed existence of scaling laws of the search behaviours of marine predators". This is an illogical conclusion to reach when their criticisms are taken as a whole. Taken together, we find their conclusion to be unsupported in light of a more balanced evaluation of the results in our papers[11,12] and the more recent literature[8,15].

An important point as a whole for progress in the burgeoning field of biological Lévy flights (and scale-invariant processes more generally) is whether there is good evidence for power laws or truncated power-laws in organism movement patterns as a result of robust analyses. Starting with studies around 2001 and through to the present time it is clear that strong evidence for naturally occurring scale-invariant spatial and temporal behaviour patterns, such as Lévy flights, is present across diverse taxa and in both natural environments and under controlled conditions[11,12,14-16,20-34]. With continued advances in animal-attached telemetry and data acquisition technology, together with the most robust statistical methods having been tested successfully with large datasets, that are now well known to a broad range of researchers entering this field of study, indicates that an important phase of work has now commenced: to understand not only when and where biological Lévy flights occur, but how and why they have might have arisen in organisms.